\documentclass[12pt]{article}

\usepackage{amssymb}
\usepackage{amsmath}
\usepackage{amscd}
\usepackage{latexsym}
\usepackage{cite}

\newcommand{\be}{\begin{equation}}
\newcommand{\ee}{\end{equation}}

\newcommand{\dlt}{\delta}

\newcommand{\bt}{\beta}

\newcommand{\al}{\alpha}
\newcommand{\ra}{\rightarrow}

\newcommand{\gm}{\gamma}

\newcommand{\cM}{{\cal M}}
\newcommand{\cA}{{\cal A}}
\newcommand{\cL}{{\cal L}}

\begin{document}

\begin{center}
{\Large{\bf Physics of risk and uncertainty\\  in quantum decision 
making} \\ [5mm]

V.I. Yukalov$^{1,2,a}$ and D. Sornette$^{1,3,b}$} \\ [3mm]

{\it $^1$Department of Management, Technology and Economics, \\
Swiss Federal Institute of Technology,
ETH Zurich, Kreuzplatz 5, Zurich 8032, Switzerland \\ [2mm]

$^2$Bogolubov Laboratory of Theoretical Physics, \\
Joint Institute for Nuclear Research, Dubna 141980, Russia \\ [2mm]

$^3$Swiss Finance Institute, University of Geneva, \\
CH-1211 Geneva 4, Switzerland}

\end{center}

\vskip 2cm

\begin{abstract}

The Quantum Decision Theory, developed recently by the authors, is 
applied to clarify the role of risk and uncertainty in decision 
making and in particular in relation to the phenomenon of dynamic 
inconsistency. By formulating this notion in precise mathematical 
terms, we distinguish three types of inconsistency: time inconsistency, 
planning paradox, and inconsistency occurring in some discounting 
effects. While time inconsistency is well accounted for in classical 
decision theory, the planning paradox is in contradiction with 
classical utility theory. It finds a natural explanation in the frame 
of the Quantum Decision Theory. Different types of discounting effects 
are analyzed and shown to enjoy a straightforward explanation within the 
suggested theory. We also introduce a general methodology based on 
self-similar approximation theory for deriving the evolution equations 
for the probabilities of future prospects. This provides a novel 
classification of possible discount factors, which include the 
previously known cases (exponential or hyperbolic discounting),
but also predicts a novel class of discount factors that decay
to a strictly positive constant for very large future time horizons. This class may be 
useful to deal with very long-term discounting situations associated 
with intergenerational public policy choices, encompassing issues such 
as global warming and nuclear waste disposal. 
\end{abstract}

\vskip 1cm

{\parindent=0pt

Communication complexity - 

89.75.-k Complex systems - 03.67.Hk Quantum communication   

\vskip 2cm

$^a$e-mail: yukalov@theor.jinr.ru

$^b$e-mail: dsornette@ethz.ch  }

\newpage

\section{Introduction}

The concept of risk is widely used in economics, finance, psychology, 
as well as in everyday life. Respectively, there exist several definitions 
of risk and different ways of evaluating it. In any application, the 
notion of risk is always related to the necessity of taking decisions under 
uncertainty. It is impossible to achieve optimal results in any science
without correct decisions, leading to optimal
consequences following from the taken decision. This is why the notion of 
risk and the problem of its evaluation has, first of all, to be understood 
in the frame of decision theory. It is precisely the aim of the present paper 
to formulate a novel approach for taking into account the risk in decision 
making and to demonstrate in concrete examples, related to temporal effects 
in making decisions, that this new approach is free of defects and paradoxes 
plaguing the application of standard decision theory.    

Classical decision theory is based on expected utility theory, which was 
advanced by Bernoulli \cite{1} and was shaped into a rigorous 
mathematical theory by von Neumann and Morgenstern \cite{2}. In this 
theory, a decision maker chooses between several lotteries, or gambles, 
each being composed of a set of outcomes, equipped with a 
probability measure. Initially \cite{2}, the probabilities were assumed 
to be objective. Savage \cite{3} extended utility theory to the case 
of subjective probabilities. Savage's generalization has been demonstrated 
to be tremendously flexible in representing the attitude of decision makers 
towards risk and uncertainty. Starting with Pratt \cite{4} and Arrow \cite{5}, 
different measures of risk have been proposed. Extensions and modern 
developments are covered, e.g., in \cite{6,7,7b}.

Notwithstanding a remarkable breadth of successful applications,
classical decision theory, when applied to real humans, leads to a variety 
of paradoxes that remain unsolved in its framework. The first such 
anomaly was described by Allais \cite{8}, which is now known as the Allais 
paradox. Other well known paradoxes are Ellsberg's paradox \cite{9}, 
Kahneman-Tversky's paradox \cite{10}, the conjunction fallacy \cite{11,12}, 
the disjunction effect \cite{13}, and Rabin's paradox \cite{14}. 
These and other paradoxes are reviewed in Refs.~\cite{15,16}. 

There has been many attempts to modify expected utility theory in order 
to get rid of the paradoxes that plague its application to the processes 
involving decision making of real human beings. One of these approaches is 
the cumulative-prospect theory or reference-point theory \cite{17}, 
which assumes that decision making is not based on the absolute 
evaluation of payoffs but depends on a reference point that is specific
to the present state of the decision maker. Because the reference 
point is shifted as a result of the consequences emerging from a first 
decision, the subsequent decision performed, according to the 
reference-point theory, is therefore sensitive to the difference between 
subsequent payoffs rather than to the absolute payoff deriving solely from 
the second decision. 

One of the main problems encountered when using 
reference-point theory is that the reference point of a decision maker is 
not uniquely defined: for a similar payoff history, each decision maker can possess (and actually does 
possess) his/her own specific reference point, which is generally unobservable. 
Moreover, reference-point theory is more suited to address those anomalies 
that arise in gambles involving at least two-steps, in which the reference point can 
be expected to be shifted after each outcome. But, the majority of 
paradoxes appear in single-step gambles, where reference-point theory is 
not applicable. In the hope of explaining the paradoxes mentioned above, 
many other variants of the so-called non-expected utility theories have 
been suggested. A review of a variety of such non-expected utility 
theories can be found in Machina \cite{18,19,20}. A rigorous analysis of 
these theories has been recently performed by Safra and Segal \cite{21}, 
who concluded that the non-expected utility theories cannot explain all 
paradoxes. Though it is possible to invent a modification of utility 
theory that will fit one or a few paradoxes, the problem is that many 
others will remain unexplained at best, or new inconsistencies will arise 
at worst.     

The basic difficulty in taking into account and evaluating risk, when 
deciding under uncertainty, is that the usual approaches assume
that decision makers are rational. However, real human beings are only partially 
rational \cite{21a}, as is well documented by numerous empirical data 
in behavioral economics and neuroeconomics \cite{21b,21c,21d,21e}. Risk 
is always related to emotions. But how could one describe emotions
within a quantitative framework suitable for decision making? 

A new approach to decision making, called Quantum Decision Theory (QDT),
 has been advanced in Refs.~\cite{15,16,22}. The main idea of this approach 
is to take into account that realistic decision-making problems are 
composite, consisting of several parts intimately interconnected, 
intricately correlated, and entangled with each other. Several intended 
actions can interfere with each other, producing effects that cannot be 
simply measured by ascribing a classical utility function. The complexity 
involved in decision making reflects the interplay between the decision 
maker's underlying emotions and feelings and his/her attitude to risk and 
uncertainty accompanying the decision making process. In order to take 
account of these subtle characteristics in the most self-consistent and 
simple way, we suggest to use the mathematical techniques based on the 
quantum theory of measurement of von Neumann \cite{23} and developed by 
other authors (see, e.g., Refs.~\cite{24,24a}). This is the reason 
for referring to this new approach under the name Quantum Decision Theory 
(QDT). It is important to stress that we do not assume that human brains 
are quantum objects. It should just be understood that we use the 
techniques of complex Hilbert spaces, as a convenient mathematical toolbox
that provides a parsimonious and efficient description of the complex 
processes involved in decision making. 

In our previous papers \cite{15,16,22}, we formulated the mathematics 
of QDT and showed that this approach provides a straightforward 
explanation of practically all known paradoxes of classical decision 
making. However, we have not yet considered the class of 
so-called dynamical inconsistencies that arise in decisions (under risks
and/or uncertainty) that compare different time horizons. The aim of 
the present paper is to analyze this class of inconsistency in the frame 
of QDT, explaining those effects that have remained unexplained in the 
standard theory. 

Our theory should not be confused with the approach that is called 
``quantum probabilities from decision theory", where one attempts to 
derive the rule of defining the probability in quantum mechanics from 
classical decision theory. To be more precise, let us recall that 
probabilities enter quantum mechanics via the Born rule, according to which 
the probability of each outcome of a measurement is prescribed by the 
squared amplitude of the corresponding term in the given quantum-mechanical 
state \cite{25}. Deutsch \cite{25a} argued that the Born rule could be 
derived from the notion of rational preferences of standard classical 
decision theory. This argument was reconsidered by Wallace \cite{25b,25c,25d}
who showed that the Deutsch way of reasoning, first, necessarily requires the 
Everett \cite{25e} many-word interpretation of quantum mechanics and, second,
needs additional assumptions that have nothing to do with classical 
decision theory. A very detailed analysis of the Deutsch-Wallace arguments 
has recently been given by Lewis \cite{25f}, who has persuasively 
demonstrated that there are several serious drawbacks in the Deutsch-Wallace 
picture. First of all, the Everett many-word interpretation has its own 
problem related to its basic assumption that, after each measurement, the 
observer branches into a number of successors living in different words. 
The number of such branches is not well defined and even can be infinite. 
According to Lewis \cite{25f}, ``the number of branches associated with an 
outcome is unknowable, undefined, and uncountable, and hence branch-counting 
rules are simply unusable". Lewis also showed that there are other gaps in 
the mathematics of Deutsch and Wallace, which invalidate the proof that
the Born rule could be derived form classical decision theory \cite{25f}.         

In our theory, we adopt the quantum-mechanical rules as its very
foundation, never trying to derive them from some other assumptions. The mathematics we 
employ is in complete agreement with the von Neumann axiomatics \cite{23}.
Using the techniques of quantum theory, we develop the 
quantum decision theory that can be applied to real alive beings.

The theory of quantum measurement considers
only passive quantum systems subject to a measurement 
procedure imposed by an external observer. A principal
difference is that our theory 
describes an active decision maker. Mathematically, an active decision maker 
is characterized by his/her own strategic states describing his/her main 
personal preferences. In contrast, in quantum measurements performed
over a passive system, there is no preferred quantum states, and any basis 
can be employed. 
 
 Moreover, our approach is completely different from the theory of quantum games,
suggested by Meyer \cite{25M} (see the review articles \cite {25E, 25F}). 
What is common for both these theories is merely the use of the quantum 
theoretical techniques, but their mathematical structure is very 
different. The general setup of a quantum game is as follows. One considers 
a passive quantum system (gamble source), several observers (players), and 
an external machine (judge). The system is prepared in a quantum state. The 
machine acts on this state entangling it. Each of the players in turn acts on 
the resulting entangled state by a unitary transformation. In this process, 
the players can exchange information between themselves and with the machine.
Then the machine again acts on the obtained state disentangling it and 
producing the final product state. The payoffs are calculated according to 
the classical rule with additive probabilities, hence, there is no 
interference in this final stage. This scheme can be considered as a variant of quantum 
computation and communication. Contrary to this scheme, in our approach, we 
consider only a single decision maker and not several ones. Of course, 
the single decision maker can represent a group of people that act as a single person.
There is no passive 
quantum system, but the decision maker represents himself/herself an active 
system acting according to quantum rules. The decision maker does not produce 
unitary transformations on the given states. There are no external judges or 
machines. Since the calculations are made by quantum rules, this involves 
nonadditive quantum probabilities and the related interference terms that are 
of crucial importance for the analysis of what constitutes the optimal 
decision. Thus, the overall structure of our theory is principally different 
from the setup of quantum games.  
  
In section 2, we provide a brief summary of the architecture of QDT
that is needed for our analysis. Section 3 dissects the three classes
of dynamic inconsistency (time inconsistency, planning paradox and 
discounting effects) and applies QDT to them. Section 4 presents
a quantitative formulation of the dynamics of prospects, in which
hyperbolic discounting is derived from simple principles. Section 5 concludes.
Let us stress once more that the dynamic effects have not been treated in 
our previous articles \cite{15,16,22}.

\section{Quantum decision theory}

In this section, we give a brief formulation of the theory to be 
used. We follow the scheme of Refs.~\cite{16,22}, employing Dirac's
notation \cite{25,26} for the states belonging to the Hilbert spaces. 
To be precise, we recall below the basic definitions and axioms of QDT.

\subsection{Main definitions \label{jtwnaa}}

{\bf Definition 1}. {\it Action ring}. The set of intended actions 
$A_n$, enumerated with an index $n$, forms an action ring
\be
\label{eq1}
\cA = \{ A_n: \; n =1,2,\ldots, N\} \; .
\ee
The ring is equipped with the binary operations, namely the addition and 
multiplication: for each $A_m$ and $A_n$ belonging to $\cA$, 
$A_m+A_n$ and $A_m A_n$ also belong to $\cA$. The addition is 
associative, so that $A_1+(A_2+A_3)=(A_1+A_2)+A_3$, and reversible, in 
the sense that $A_1+A_2=A_3$ yields $A_1=A_3-A_2$. The multiplication 
is distributive, $A_1(A_2+A_3)=A_1 A_2+A_1 A_3$, and idempotent, 
$A_n A_n=A_n^2= A_n$. But, generally, it is not commutative, so that 
$A_mA_n$ does not necessarily equal $A_nA_m$ when $m$ and $n$
are different. There exists an empty action, such that $A_n0=0A_n=0$. 
Two actions $A_m$ and $A_n$ are disjoint when $A_mA_n=A_nA_m=0$.

\vskip 2mm

{\bf Definition 2}. {\it Action modes}. The elements of the action 
ring, the actions, can be composite  
\be
\label{eq2}
A_n = \bigcup_{\mu=1}^{M_n} A_{n\mu} \qquad (M_n > 1) \; ,
\ee                                                        
being composed of several representations, called modes, labelled by 
$\mu$. Different modes are assumed to be disjoint, 
$$
A_{n\mu} A_{n\nu} = \delta_{\mu\nu} A_{n\mu} \; ,
$$ 
where $\delta_{\mu\nu}$ is the Kronecker delta.
An action is composite if $M_n >1$, in the other case, it is simple.  

\vskip 2mm

{\bf Definition 3}. {\it Action prospects}. A more complex structure 
is an action prospect

\be
\label{eq3}
\pi_j = \bigcap_{n} A_{j_n} \qquad (A_{j_n}\in\cA) \; ,
\ee                                               
which is a conjunction of several actions. The prospect is composite 
if it includes composite actions, while it is simple if all actions in 
(\ref{eq3}) are simple. Generally, the number of factors in the 
intersection can be different, depending on the definition of the 
prospects. Since the products of actions pertaining to the ring, by the 
ring structure, also pertain to the ring, then the prospects are also members 
of the same ring.  

\vskip 2mm

{\bf Definition 4}. {\it Elementary prospects}. A prospect is called 
elementary if all actions in its definition (\ref{eq3}) are simple, being 
represented by single modes. The elementary prospects 
\be
\label{eq4}
e_\al =  \bigcap_{n=1}^N A_{i_n \mu_n}
\ee                                                            
are labelled by the binary multi-index 
$$
\al = \{ i_n, \mu_n : \; n=1, 2, \ldots, N \}_\al \; .
$$
The set $\{\al\}$ has cardinality ${\rm card}\{\al\}=\prod_{n=1}^N
M_n$. All elementary prospects are disjoint with respect to each 
other, 
$$
e_\al e_\bt = \dlt_{\al \bt} e_\al \; .
$$
Here and in what follows, the cardinality $N$ is the same as in Def. 1.

\vskip 2mm

{\bf Definition 5}. {\it Prospect lattice}. A particular family of 
prospects composes a prospect lattice 
\be
\label{eq5}
\cL = \{ \pi_j: \; j = 1,2, \ldots, N_L \} \; ,
\ee                                                              
where the binary operations $\geq$ and $\leq$ are assumed to be defined, 
ordering the prospects so that, for each  pair $\pi_i$ and $\pi_j$, 
either $\pi_i\geq\pi_j$ or $\pi_i\leq\pi_j$. For a while, it is sufficient 
to keep in mind that the prospects can be ordered. The explicit ordering 
procedure will be prescribed below in Def. 16.

\vskip 2mm

{\bf Definition 6}. {\it Mode states}. To each mode $A_{n\mu}$ there 
corresponds a complex function
\be
\label{eq6}
| A_{n\mu} > : \; \cA \; \ra \; \mathbb{C}          
\ee
called the mode state. The fact that each mode is idempotent and 
different modes are disjoint is expressed through the orthonormality 
condition for the scalar product $<A_{n\mu}|A_{n\nu}>=\dlt_{\mu\nu}$.

\vskip 2mm

{\bf Definition 7}. {\it Mode space}. The closed linear envelope
\be
\label{eq7}
\cM_n =  {\rm Span} \{ | A_{n\mu} >\; : \; 
\mu =1,2,\ldots,M_n  \} \; ,
\ee                                                      
spanning all mode states, equipped with a scalar product, 
is the mode space. This is a Hilbert space of dimensionality 
${\rm dim}\cM_n=M_n$.

\vskip 2mm

{\bf Definition 8}. {\it Basic states}. To each elementary prospect 
(\ref{eq4}), there corresponds a complex function 
\be
\label{eq8}
|e_\al > \; = \; | A_{i_1\mu_1} A_{i_2\mu_2} \ldots 
A_{i_N\mu_N} > \; = \; \bigotimes_{n=1}^N |A_{i_n\mu_n} >  \; ,                              
\ee
called a basic state. Since an elementary prospect (\ref{eq4}) is 
a conjunction of single modes, and different modes are disjoint with 
each other, this is expressed as the orthonormality condition for 
the scalar product $<e_\al|e_\bt>=\dlt_{\al\bt}$.      

\vskip 2mm

{\bf Definition 9}. {\it Mind space}. The closed linear envelope
\be
\label{eq9}
\cM = {\rm Span} \{ |e_\al>\; : \; \al \in \{ \al\} \; \} =
\bigotimes_{n=1}^N \cM_n \; ,                                           
\ee
spanning all basic states, endowed with a scalar product, is the 
mind space. This is a Hilbert space of dimensionality 
${\rm dim}\cM=\prod_{n=1}^N M_n$.

\vskip 2mm

{\bf Definition 10}. {\it Prospect states}. To each prospect 
(\ref{eq3}), there corresponds a complex function $|\pi_j>$ 
belonging to the mind space $\cM$. Because the prospects, 
generally, are composite, they are not necessarily normalized 
and orthogonal to each other. 

\vskip 2mm

{\bf Definition 11}. {\it Strategic states}. In the mind space 
(\ref{eq9}), there exist fixed reference states  $|\psi_s>\;\in\cM$, 
which characterize the features typical of a given decision maker. 
These states are orthonormal, such that $<\psi_s|\psi_{s'}>=\dlt_{s s'}$.
But they do not necessarily form a basis. The existence of the 
strategic states is the principal point distinguishing QDT from the 
usual theory of quantum measurements.

\vskip 2mm

{\bf Definition 12}. {\it Mind strategy}. The collection of all 
strategic states $|\psi_s>$, equipped  with their weights $w_s$, 
forms the mind strategy
\be
\label{eq10}
\Sigma = \{ | \psi_s> , w_s : \; s= 1,2, \ldots, S \}  \; ,                 
\ee
where 
\be
\label{eq11}
\sum_{s=1}^S w_s = 1 \; , \qquad 0 \leq w_s \leq 1  \; .
\ee
The mind strategy describes the decision-maker character, his/her 
main beliefs and principles, according to which he/she makes decisions.

\vskip 2mm

{\bf Definition 13}. {\it Prospect operators}. Each prospect state 
$|\pi_j>$ defines the prospect operator
\be
\label{eq12}
\hat P (\pi_j) =  |\pi_j > < \pi_j |  \; ,                                 
\ee
where $<\pi_j|$ is the Hermitian conjugate to $|\pi_j>$. The prospect 
operators, by definition, are self-adjoint. The family of all 
prospect operators forms the involutive bijective algebra
$$
{\cal P} = \{ \hat P(\pi_j) : \; \pi_j \in \cL \} \; .
$$
This algebra is analogous to the algebra of local observables in 
quantum theory. 

\vskip2mm

{\bf Definition 14}. {\it Operator averages}. The average of a prospect 
operator (\ref{eq12}) is the sum
\be
\label{eq13}
 < \hat P(\pi_j) > \; = \; 
\sum_{s=1}^S w_s \; < \psi_s |\hat P(\pi_j) | \psi_s >                      
\ee
of its matrix elements over the strategic states.

\vskip 2mm

{\bf Definition 15}. {\it Prospect probability}. The probability of 
a prospect $\pi_j\in\cL$ is the average 
\be
\label{eq14}
p(\pi_j) \; = \; < \hat P(\pi_j) >
\ee                                                                      
of the prospect operator (\ref{eq12}), with the normalization condition
\be
\label{eq15}
\sum_{j=1}^{N_L} p(\pi_j) = 1 \; ,
\ee                                                          
where the summation is over the whole prospect lattice $\cL$.

\vskip 2mm

{\bf Definition 16}. {\it Prospect ordering}. A prospect $\pi_1$ is 
indifferent to a prospect $\pi_2$ if and only if their probabilities 
coincide,
\be
\label{eq16}
p(\pi_1) =  p(\pi_2) \qquad (\pi_1 = \pi_2) \; .
\ee   
And a prospect $\pi_1$ is preferred to $\pi_2$ if and only if 
\be
\label{eq17}
p(\pi_1) >  p(\pi_2) \qquad (\pi_1 > \pi_2) \; .
\ee                            
The ordering of prospects through the relation between their 
probabilities defines the explicit ordering in the prospect lattice 
(\ref{eq5}). The prospect $\pi^*$ with the largest probability 
$p(\pi^*)=\sup_j p(\pi_j)$ is called optimal.

\vskip 2mm

{\bf Definition 17}. {\it Partial probabilities}. The probability 
\be
\label{eq18}
p(\pi_j e_\al) \; = \; 
< \hat P(e_\al) \hat P(\pi_j) \hat P(e_\al) >
\ee                                                                        
of a conjunction prospect $\pi_j e_\al$ defines the partial 
probability of realizing an elementary prospect $e_\al$ when deciding 
on the prospect $\pi_j$. The partial probabilities are normalized as 
\be
\label{eq19}
\sum_{j,\al} p(\pi_j e_\al) = 1 \; ,
\ee                        
where the sum is over all $\pi_j\in\cL$ and all $e_\al$.

\vskip 2mm

{\bf Definition 18}. {\it Attraction factor}. The variable
\be
\label{eq20}
q(\pi_j) = \sum_{\al\neq \bt} < \hat P(e_\al) 
\hat P(\pi_j) \hat P(e_\bt) >
\ee
quantifies the attractiveness of the prospect $\pi_j$ for a decision  maker 
with respect to risk, uncertainty, and biases. It arises due to
the interference between the intended actions of
a given prospect $\pi_j$, which occurs during the decision process.

\vskip 2mm

{\bf Definition 19}. {\it Attraction ordering}. The prospects are 
ordered with respect to their attractiveness for a decision maker. 
A prospect $\pi_1$ is more attractive than a prospect $\pi_2$ if and 
only if   
\be
\label{eq21}
q(\pi_1) >  q(\pi_2) \; .
\ee 
The prospects $\pi_1$ and $\pi_2$ are equally attractive if and only 
if
\be
\label{eq22}
q(\pi_1) =  q(\pi_2) \; .
\ee 
The impact in decision making
of emotions and feelings, which are 
known to be important and practically inseparable from logical 
deliberation \cite{27}, are quantified by the attraction factor.
The ordering of prospects with respect to their attractiveness,
quantified by the attraction factor (\ref{eq20}), is a principal 
ingredient of QDT.

\vskip 2mm

{\bf Definition 20}. {\it Attraction conditions}. The distinction 
between more or less attractive prospects is formalized by the 
following rule. A prospect $\pi_1$ is more attractive than a prospect 
$\pi_2$, when it is connected with:

\vskip 2mm
(a) more certain gain,
   
(b) less certain loss,
   
(c) higher activity under certainty,
   
(d) lower activity under uncertainty. 

\vskip 2mm

These characteristics describe the aversion of a decision maker to 
risk, uncertainty, and presumed loss. 

\subsection{A few theorems}

The above definitions constitute the basis of QDT \cite{15,16,22}. 
They allow us to derive the following theorems proved in Ref.~\cite{16}, 
which will be needed below.

\vskip 2mm

{\bf Proposition 1} ({\it Prospect probability}). The probability of 
a prospect $\pi_j\in\cL$ is 
\be
\label{eq23}
p(\pi_j) = \sum_{\al} p(\pi_j e_\al) \; + \; q(\pi_j) \; ,
\ee
where the summation is over the elementary prospects $e_\al$.

\vskip 2mm

{\bf Proposition 2} ({\it Attraction alternation}). The sum of all 
attraction factors (\ref{eq20}) is equal to zero:
\be
\label{eq24}
\sum_{j=1}^{N_L} q(\pi_j) =  0\; ,
\ee
where the summation is performed over all $\pi_j\in\cL$.

\vskip 2mm

{\bf Proposition 3} ({\it Preference criterion}). A prospect 
$\pi_1\in\cL$ is preferred to a prospect $\pi_2\in\cL$ if and only 
if 
\be
\label{eq25}
\sum_\al \; [ p(\pi_1 e_\al) - p(\pi_2 e_\al) ] \; > \; 
q(\pi_2) - q(\pi_1) \; .
\ee 

\vskip 2mm

{\bf Remark}. From the form of prospect probability (\ref{eq23}), 
together with condition (\ref{eq19}) and property (\ref{eq24}), it 
is immediately seen that the normalization condition (\ref{eq15}) is 
always valid.

\vskip 2mm 

These theorems imply that the probability of taking a given
decision is controlled by the levels of attraction of the
different competing prospects, thus emphasizing the emotional 
component of the decision process. Indeed, the choice of a specific 
prospect among several alternatives depends not solely on its value 
given by the first term in the right-hand side of Eq. (\ref{eq23}), 
but also on its attractiveness quantified by the attraction factor 
(\ref{eq20}). In classical decision theory, only values measured by 
a utility function are considered, but emotions and feelings are not 
taken into account. In QDT, the later are embodied in the new 
ingredients, the attraction factors.

Two essential characteristics distinguish QDT from classical
utility theory:

\vskip 2mm

(i) QDT is a probabilistic theory, in which each prospect is 
associated with its probability, which has a subjective component 
captured by the attraction factor. The prospect probability 
can be measured experimentally, by interpreting it as a relative 
frequency, that is, it corresponds to the relative ratio of decision 
makers accepting the given prospect. This probabilistic framework 
accounts for the observations that, under the same conditions, 
different people endowed with a priori the same preferences 
may make different decisions. In contrast, classical 
utility theory is deterministic, with its prescription to the decision 
maker forcing him/her to accept the unique alternative which 
corresponds to the maximal expected utility.

\vskip 2mm

(ii) In addition to the payoff values, QDT takes also into account 
the attractiveness of the analyzed prospects, quantified by their 
attraction factors (\ref{eq20}). These attraction factors are absent 
in utility theory. Therefore, a partial reduction of QDT to classical 
decision theory is obtained by setting the attraction factor to zero.

\vskip 2mm

\subsection{Binary mind}

To make the structure of the theory clearer, it is instructive 
to consider the particular case of a binary mind. This case
is also of intrinsic interest because
the majority of paradoxes can be treated and explained 
in this specific frame.

The binary mind corresponds to considering only two actions, while 
each of them can possess a number of representation modes. Let these 
actions be
\be
\label{eq26}
A = \bigcup_{j=1}^{M_1} A_j \; , \qquad
B = \bigcup_{\mu=1}^{M_2} B_\mu \; .
\ee  
Hence, there are two mode spaces
\be
\label{eq27}
\cM_1  = {\rm Span} \{ | A_j >\; : j = 1,2 \ldots, M_1 \} \; , 
\qquad
\cM_2  = {\rm Span} \{ | B_\mu >\; : \mu = 1,2 \ldots, M_2 \} \; .
\ee 
The mind space is the tensor product of these two mode spaces
\be
\label{eq28}
\cM = \cM_1 \otimes \cM_2 \; ,
\ee
hence its name ``binary''. This should not be confused with the
dimensionality ${\rm dim}\cM=M_1 M_2$ of the binary mind,
which can be large.

The elementary prospects (\ref{eq4}) are $e_{j\mu}= A_j B_\mu$, 
and the basic states (\ref{eq8}) become
\be
\label{29}
| e_{j\mu} > \; = \; | A_j B_\mu> \; \equiv \; | A_j> \; \otimes \;
|B_\mu > \; .
\ee
The action prospects (\ref{eq3}) can be constructed as $\pi_j=A_j B$, 
and the conjunction  prospects as $\pi_j e_{j\mu}=A_j B_\mu$. 
According to Eq. (\ref{eq23}), the prospect probabilities are    
\be
\label{eq30}
p(\pi_j) = \sum_{\mu=1}^{M_2} p( A_j B_\mu) \; + 
\; q(\pi_j) \; .
\ee

One can draw the following
analogies between the quantities of QDT presented above, and 
those of classical utility theory. The set $B$ of 
modes $B_\mu$ corresponds to the set of payoffs. Complementing this 
set by the related weights $p_j(B_\mu)$ defines a lottery $L_j$. 
The weights $p_j(B_\mu)$ can be expressed in terms of the 
conditional probabilities: $p_j(B_\mu)= p(A_j|B_\mu)$. This defines
the probability of getting payoff $B_\mu$ 
in lottery $L_j$. The analog of the expected utility is the sum
\be
\label{egwerg}
\sum_{\mu=1}^{M_2} p(A_j B_\mu) = 
\sum_{\mu=1}^{M_2} p(A_j|B_\mu) p(B_\mu) \; ,
\ee
where $p(B_\mu)$ is a normalized measure of the payoff $B_\mu$.

\vskip 2mm

With QDT, it is possible to explain all paradoxes emerging in
classical decision making \cite{15,16}. To give an idea how this is 
done, we present here a brief account of the resolution of Allais'
paradox \cite{8}. Allais' paradox can be described with a binary mind, 
as defined above. For the sake of brevity, we survey only the mathematical 
structure of this paradox, omitting the interpretations related to 
psychological features (see Refs.~\cite{15,16} for in-depth analysis).
A detailed description of the mathematical structure of the Allais paradox 
can be found in Ref.~\cite{16}.
\vskip 2mm

One considers two actions
as in Eq. (\ref{eq26}), with $M_1=4$ and $M_2=3$ and the
mind dimensionality ${\rm dim}\cM=M_1 M_2 = 12$. The experiment, 
demonstrating Allais' paradox, is organized in such a way that the 
balance condition
\be
\label{eq32}
p(A_1 B_\mu) + p(A_3 B_\mu) = p(A_2 B_\mu) + p(A_4 B_\mu)
\ee
holds for all $\mu=1,2,3$. The goal is to compare the prospects 
$\pi_j=A_j B$ for different $j$. Allais' paradox is that most human 
decision makers prefer the prospect $\pi_1$ to $\pi_2$, and $\pi_3$ to 
$\pi_4$ which, due to the balance condition (\ref{eq32}), leads to a 
contradiction. The fact that $\pi_1$ is preferred to $\pi_2$ translates 
in the language of QDT into the inequality $p(\pi_1)>p(\pi_2)$. The fact 
that prospect $\pi_1$ looks more attractive (less uncertain, less 
risky) than $\pi_2$ implies that $q(\pi_1)>q(\pi_2)$.  Using
(\ref{eq30}), this leads to
\be
\label{eq33}
\sum_{\mu=1}^3 \; [ p(A_2 B_\mu) - p(A_1 B_\mu) ] \; < \;
q(\pi_1) - q(\pi_2) \; .
\ee
The fact that $\pi_3$ is preferred to $\pi_4$ translates in the 
language of QDT into $p(\pi_3)>p(\pi_4)$. The larger attraction of 
$\pi_3$, compared with $\pi_4$, implies that $q(\pi_3)>q(\pi_4)$. Again 
using (\ref{eq30}), this gives
\be
\label{eq34}
\sum_{\mu=1}^3 \; [ p(A_3 B_\mu) - p(A_4 B_\mu) ] \; > \;
q(\pi_4) - q(\pi_3) \; .
\ee
Then, using the definitions of subsection 2.1 and Proposition 2 on the 
property of attraction 
alternation, invoking the balance condition (\ref{eq32}), and combining 
inequalities (\ref{eq33}) and (\ref{eq34}), we get
\be
\label{eq35}
- | q(\pi_3) - q(\pi_4) | \; < \; \sum_{\mu=1}^3 \; 
[ p(A_2 B_\mu) - p(A_1 B_\mu) \; < | q(\pi_1) - q(\pi_2) | \; .
\ee
Classical decision theory corresponds to the limit of zero 
attraction factors ($q(\pi_1)=q(\pi_2)=q(\pi_3)=q(\pi_4)=0$). In this 
case, the two inequalities (\ref{eq35}) result in a contradiction, 
since the sum in the middle cannot be larger than zero and, at the 
same time, smaller than zero. Within QDT, this contradiction does not 
arise. Actually, within QDT, Allais' paradox is explained from the 
interplay between the attraction factors of different prospects.

\section{Dynamic inconsistency}

We now use the framework of QDT to study dynamic inconsistency, 
which has not been treated in our previous articles. In economics, 
time inconsistency refers, roughly speaking, to a situation when the 
preference of a decision-maker changes over time, in such a way that 
what is preferred at one point in time is inconsistent with what is 
preferred at another point in time. In fact, there are numerous variants 
of dynamic inconsistency. By being precise, one can distinguish three 
broad classes of dynamic inconsistency: (i) time inconsistency, 
(ii) planning paradox, and (iii) discounting effects. We now examine 
each one in turn.

\subsection{Time inconsistency}

Time inconsistency is well epitomized by the Strotz's phrase \cite{28}: 
``the optimal plan of the present moment is generally one which is not 
obeyed, or that the individual's future behavior will be inconsistent 
with his optimal plan''. Various examples of this inconsistency have 
been described in the literature \cite{29,30,31}. Kydland and Prescott 
\cite{30} went so far as saying that the rational choice for future 
times ``is not an appropriate tool for economic planning'' and that 
``the application of optimal control theory is equally absurd''. 

The origin of time inconsistency is rather straightforward.
When an individual makes a plan for the far future, 
he/she cannot be conscious of all the detailed circumstances that 
will arise in that future. New information is likely to appear and, 
in addition, the already available information may be open for 
re-evaluation. Since the future situation is likely to be different, 
it will require making a decision that is likely to differ from the 
current decision. The current decision for the future action then 
turns out to be sub-optimal when the future becomes the present.

There is no real paradox in this time inconsistency and its solution 
can be readily obtained: when making a decision for the distant future, 
it is necessary to try to predict future changes and include these 
forecasts in the decision making process. This recipe was suggested 
for instance by Strotz \cite{28} who gave, as an example, the behavior 
of Odysseus  when his ship was approaching the Sirens. Wishing to 
hear the Sirens' songs (short-term gratification) but mindful of the 
possible delayed danger (falling prey to the sirens), he ordered his 
men to close their ears with beeswax and to bind him to the mast of 
the ship. He also ordered his men not to heed his cries while they 
would pass the Sirens. In that way, Odysseus limited his future 
agency and binded himself to a restriction (to the mast) to survive 
the long-term consequences of his decision. Other numerous example 
are known, related to pension savings, health insurance, and so on. 
When making plans for the far future, one tries to anticipate the 
obstacles that may arise and one imposes restrictions and commitments 
that oppose the change of decision that would result otherwise due to 
time inconsistency. With the imposed commitments, time inconsistency
disappears and the present decision becomes the optimized one for the 
future state. As experiments show \cite{32}, even rats possess the 
ability of making decisions that take into account an estimation of 
future events. We conclude that both the origin and the solution for 
time inconsistency are well understood and do not require invoking 
additional concepts for their interpretation.

\subsection{Planning paradox}

Consider a situation in which an individual makes a plan for a short future 
period of time, such that no novel information will become available and the 
individual himself/herself does not change over that period. In the absence 
of any new information and of any change, the decision should be unchangeable 
as well. The invariance of the decision in that sense is referred to as the 
principle of dynamic consistency in classical decision theory. 

However, it often happens that the decision maker does change the plan, 
for not apparent reason. A stylized example of this type of planning paradox
is a smoker who plans to stop smoking tomorrow, while enjoying the pleasure 
of smoking today. Making this plan, he/she promises to stop smoking, 
understanding well that he/she will forgo future pleasures, for the 
anticipation of higher health benefits. The next day, while the plan and the 
utility resulting from its consequences have not changed, it is often 
observed that the human beings change their plan, and continue smoking. 

\subsubsection{Mathematical formulation of planning paradox and its resolution
\label{bgwgbw}}

Such a behavior poses a real paradox within expected utility theory. Let us 
formulate this paradox in precise mathematical terms. When deciding to stop 
smoking in a plan, one keeps in mind the following intended actions: 
\begin{itemize}
\item planning to stop smoking  tomorrow ($A_1$),  
\item planning to continue smoking tomorrow ($A_2$),
\item wishing to have good health ($B_1$),
\item paying little attention to health ($B_2$).
\end{itemize}

The decision to stop smoking in reality corresponds to the following 
intended actions:
\begin{itemize}
\item stop smoking in reality ($A_3$),
\item continue smoking in reality ($A_4$),
\item wishing to have good health ($B_1$),
\item paying little attention to health ($B_2$).
\end{itemize}

The related four action sets are $X_j=\{ A_j B_\mu:\;\mu=1,2\}$, with 
$j=1,2,3,4$. Following utility theory, and ascribing probabilities to these 
actions, one gets the corresponding lotteries $L_j=L_j(X_j)$. Note that 
the utility functions of the actions $A_1 B$ and $A_3 B$, where $B=B_1+B_2$, 
are the same when expressed for tomorrow, since the two actions of stopping 
smoking become equivalent. Similarly, the utility functions tomorrow of 
the actions $A_2 B$ and $A_4 B$ are equal, since continuing smoking is the 
same action, with the same consequences. Therefore, the expected utilities 
of the lotteries $L_1$ and $L_3$ are equal: $U(L_1)=U(L_3)$. And analogously, 
$U(L_2)=U(L_4)$. But many individuals prefer $L_1$ to $L_3$,  which implies 
that, for these individuals, $U(L_1)>U(L_3)$. The same individuals also
choose $L_4$ over $L_2$, which implies that $U(L_2)<U(L_4)$. This leads to a 
contradiction violating the principle of dynamic consistency of classical 
decision making. 

Let us now show how this paradox can be explained within QDT. 
As above, we need to consider the intended actions $A_j$, with $j = 1,2,3,4$
and the set  $\{ B_\mu\}$. In addition, the decision of stopping smoking in 
reality is accompanied by the following intended actions:
\begin{itemize}
\item getting pleasure from smoking ($C_1$),
\item having no pleasure from smoking ($C_2$), 
\item agreeing to suffer because of addiction ($D_1$),
\item refusing to suffer from addiction ($D_2$).
\end{itemize}
These additional intentions reflect emotional feelings of the decision maker, 
which are not taken into account in classical utility theory.

The prospects that need now to be compared are 
\be
\label{eq36}
\pi_1 = A_1 B, \qquad \pi_2 = A_2 B \qquad 
\left ( B = \bigcup_\mu B_\mu \right ) \; 
\ee
and 
\be
\label{eq37}
\pi_3 = A_3 B C D, \qquad \pi_4 = A_4 B C D,
\ee
where
$$
B = B_1 + B_2, \qquad C = C_1 + C_2, \qquad D = D_1 + D_2.
$$

The value of quitting smoking, either today or tomorrow, has the same 
determined value. Respectively, the value of continuing smoking is 
also determined, being the same either today or tomorrow. In both these 
cases, the utility of stopping smoking is larger than that of continuing 
smoking, which can be expressed as the inequality
\be
\label{eq38}
\sum_\mu \; p(A_1 B_\mu) > \sum_\mu \; p(A_2 B_\mu) \; .
\ee
In QDT, the attraction factors are taken into account, which model the 
subjective emotions associated with different actions. Since the health 
benefits are evident, stopping smoking in a plan seems to be more attractive 
than to continue smoking. This looks easy, since the associated pain is 
not yet felt but the risk for health, associated with the continuation of 
smoking, seems evident. This is why to stop smoking in a plan is more 
attractive than to continue smoking. Then the corresponding attraction 
factors obey the inequality $q(\pi_1 ) > q(\pi_2)$. In contrast, continuing 
smoking unconditionally amounts to abandon oneself to the pleasure of addiction, 
which is preferred in general to the failure of not abiding to a plan to 
abandon smoking, given that the health benefits are felt to be uncertain. 
One can summarize these emotions by saying that continuing smoking in reality 
is more attractive than stopping smoking. This is formulated mathematically 
by the inequality $q(\pi_4 ) > q(\pi_3)$ for the corresponding attraction 
factors. Summarizing, we have
\be
\label{eq39}
q(\pi_1 ) > q(\pi_2) \; , \qquad q(\pi_4) > q(\pi_3) \; .
\ee
Writing the prospect probabilities according to Eq. (\ref{eq30}), and
taking into account the above discussion, shows that, in reality, the 
probability of continuing smoking becomes larger than that of stopping 
smoking when 
\be
\label{eq40}
\sum_\mu [ p(A_1 B_\mu) - p(A_2 B_\mu) ] \; < \; 
q(\pi_4) - q(\pi_3) \; .
\ee
Then it is implied immediately that the prospect $\pi_1$ is preferred to 
$\pi_2$, while $\pi_4$ is preferred to $\pi_3$. As for other paradoxes,
the absence of contradiction in QDT results from the existence of the 
attraction factors, which are absent in classical utility theory. We have 
shown that the attraction factors derive intrinsically from the Hilbert 
space structure of the theory that accounts for interference between 
prospects. Putting the attraction factors to zero recovers the inconsistency 
associated with the planning paradox. As QDT is a probabilistic theory, 
the above conclusion that  $p(\pi_1)>p(\pi_2)$ and $p(\pi_4)>p(\pi_3)$ 
does not mean that no individual can stop smoking. The general subjective
preferences embodied in the attraction factors only tell us that the majority 
of them will not be able to quit smoking. 

\subsubsection{Generalization to two-step games}

To show that the explanation proposed by QDT is general, let us 
consider another example of the planning paradox, with two-step 
gambles. In two-step gambles, decision makers are typically confronted 
sequentially with two successive gambles, with probabilities 1/2 to 
gain or to loose in each of them. Before playing the first gamble, 
participants are asked to make a planned choice as whether they would 
take the second gamble, provided the first one is either won or lost. 
Then the first gamble is played. After experiencing the actual results 
of the first gamble, decision makers are asked to make a final choice 
regarding the second gamble, whether they accept it or not. 

A number of experiments have been performed to test the dynamic 
consistency in the frame of such two-step gambles \cite{33,34,35}. 
The experiments showed that the final choices of the participants  
were frequently inconsistent with their plans, even when the 
anticipated and experienced outcomes were identical. These 
inconsistencies are found to occur in a systematic direction: 
anticipating a gain in the first gamble, decision makers planned to 
take the second gamble - but after experiencing the gain, some of them 
changed their minds and rejected the second gamble. And, anticipating 
a loss in the first gamble, the participants planned to restrain from 
the second gamble - however, experiencing the actual loss, they often 
changed their plans and accepted the second gamble. Attempts were made 
\cite{33,35} to explain this inconsistency within the framework of the 
reference-point theory \cite{17}, arguing that, after the first gamble, 
the reference point of the decision makers has been shifted. In the 
introduction Section 1, we have already discussed the weakness of the 
reference-point approach. These are the ambiguity in defining both the 
reference point as well as the shift. And, what is more important, 
the reference-point theory can be applied only to two-step or multi-step 
gambles. It is not applicable to single-step gambles. But there are 
numerous cases where the planning paradox occurs in single-step gambles,
such as in the above example of the smokers planning to stop 
smoking. In an earlier publication \cite{36}, the authors mentioned 
that the planning paradox in two-step gambles could be related to 
quantum effects. Below, we provide a concrete proof in the frame of 
QDT, by showing how the planning paradox in two-step games finds 
a natural resolution. 

The mathematical structure of the two-step gambles of the type described 
in Refs.~\cite{33,34,35} can be reduced to a structure that is similar 
to, though slightly more complicated than, the structure underlying the 
case described in the previous subsection \ref{bgwgbw}. The two-step game
proceeds as follows. The first gamble is obligatory and cannot be refused 
while the second gamble can be rejected. Specifically, the following 
alternatives are offered to the decision maker. 
\begin{itemize}
\item Assuming an anticipated gain ($C_1$) 
or loss ($C_2$) in the first gamble, the second gamble can be 
accepted ($A_1$) or rejected ($A_2$), with the chances of winning 
($B_1$) or loosing ($B_2$) being equal. 
\item After experiencing a realized gain ($C_3$) or an actual loss 
($C_4$) in the first gamble, the second gamble can be accepted ($A_1$) 
or rejected ($A_2$), with the chances of winning ($B_1$) or to loose 
($B_2$).
\end{itemize}
The planning stage, before playing the first game, is characterized by
the four prospects
\be
\label{eq41}   
\pi_1 = A_1 B C_1 \; , \qquad   \pi_2 = A_2 B C_1 \; , \qquad
\pi_3 = A_1 B C_2 \; , \qquad   \pi_4 = A_2 B C_2 \; ,
\ee
where $B=B_1+B_2$.  After having played the first game,
the decision maker faces the four new prospects
\be
\label{42}
\pi_5 = A_1 B C_3 \; , \qquad  \pi_6 = A_2 B C_3 \; , \qquad
\pi_7 = A_1 B C_4 \; , \qquad \pi_8 = A_2 B C_4 \; .
\ee
The four prospects in the planning stage form two binary lattices:
\be
\label{eq43}
\cL_1 = \{ \pi_1,\; \pi_2 \} \; , \qquad 
\cL_2 = \{ \pi_3, \; \pi_4 \} \; .
\ee
The four prospects available after playing the first game form
the two other binary lattices
\be
\label{eq44}
\cL_3 = \{ \pi_5,\; \pi_6 \} \; , \qquad 
\cL_4 = \{ \pi_7, \; \pi_8 \} \; .
\ee

Analogously to conditions in subsec. 3.2.1, it is assumed 
that the utility of accepting or rejecting the second gamble does not 
depend on whether the first gamble is assumed to be won or lost in the 
planning stage or actually won or lost in reality. This means that
\be
\label{eq45}
\sum_\mu p(A_1 B_\mu C_1 )  =  \sum_\mu p(A_1 B_\mu C_3 ) \; , \qquad 
\sum_\mu p(A_1 B_\mu C_2 )  =  \sum_\mu p(A_1 B_\mu C_4 ) \; .
\ee
Next, we model the subjective beliefs and emotions commonly observed in 
humans by specifying the attraction factors of each prospect. Many human 
beings share the gambler's fallacy \cite{Gamblerfallacy}, in which an 
observed deviation from an expected fair chance of winning or losing is 
expected to be followed by a reversal. In other words, playing a gamble 
with equal chances to win or to loose, humans often expect that, after 
winning one gamble, the chance to win a second gamble is reduced. 
Reciprocally, after loosing one gamble, the odds to win the next gamble 
are felt to increase. One can say that, after winning a  gamble, a fear to 
loose the next gamble appears. However, this fear is less intense in 
imagination than in reality. That is, the perceived risk in the planning 
stage is weaker than after the realized gain of the first game, since an 
imaginary gain or loss is less certain than the real one. This makes 
the prospect $\pi_1$ of accepting the second gamble, after an anticipated 
gain in the first gamble, more attractive than the prospect $\pi_5$ of 
really accepting the second gamble after an actual gain in the first 
gamble. This translates into
\be
\label{eq46}
q(\pi_1) > q(\pi_5) \; .
\ee
Similarly, after loosing in the first gamble, the expectation to
win in the second gamble increases, but less in imagination than 
following a realized win, hence
\be
\label{eq47}
q(\pi_3) < q(\pi_7) \; . 
\ee
We thus obtain the probabilities of the prospects $\pi_1$ and $\pi_5$ as
\be
\label{eq48}
p(\pi_1) = p(A_1 B_1 C_1 ) +  p(A_1 B_2 C_1 ) + q(\pi_1) \; ,
\qquad
p(\pi_5) = p(A_1 B_1 C_3 ) +  p(A_1 B_2 C_3 ) + q(\pi_5) \; .
\ee
Similarly, the probabilities of the prospects $\pi_3$ and $\pi_7$ are
\be
\label{eq49}
p(\pi_3) = p(A_1 B_1 C_2 ) +  p(A_1 B_2 C_2 ) + q(\pi_3) \; ,
\qquad
p(\pi_7) = p(A_1 B_1 C_4 ) +  p(A_1 B_2 C_4 ) + q(\pi_7) \; .
\ee
Comparing these probabilities, with taking account of conditions 
(\ref{eq45}), 
we get
\be
\label{eq50}
p(\pi_1) - p(\pi_5) = q(\pi_1) - q(\pi_5) \; , \qquad
p(\pi_7) - p(\pi_3) = q(\pi_7) - q(\pi_3) \; .
\ee
From Eq. (\ref{eq46}), we obtain $p(\pi_1)>p(\pi_5)$, that is, 
the first prospect is preferred to the fifth prospect, $\pi_1>\pi_5$:
individuals choose to play the second game more often when 
they do not know the outcome of the first game but expect a gain,
than after the gain is realized. From Eq. (\ref{eq47}), we see that $p(\pi_7)>p(\pi_3)$, 
hence the seventh prospect is preferred to the third one, $\pi_7>\pi_3$:
individuals choose more often to play the second game after
losing the first game than when imagining that they could lose 
before playing the first game. Thus, no contradiction arises within QDT.           

We again emphasize that the preference for one prospect at the expense
of a second prospect
does not imply that all decision makers choose it, but only that
the fraction of decision makers preferring that
prospect is larger than the fraction of decision makers choosing the second
prospect. Depending on the gain prizes and on the loss amounts,
the resulting differences between the corresponding prospect 
probabilities may be small. For example, in the experiment of Barkan and 
Busemeyer \cite{35} on the planning paradox, the probabilities, measured 
as the average fractions of decision makers taking the corresponding 
alternatives are as follows. In the planning stage before playing the
first game, one has
\be
p(\pi_1) = 0.60 \; , \qquad p(\pi_2) = 0.40 \; ,
\qquad  p(\pi_3) = 0.63\; , \qquad p(\pi_4) = 0.37 \; .       
\ee
After the gain or loss of playing the first game are known, the 
probabilities of the different prospects are
\be
p(\pi_5) = 0.53\; , \qquad p(\pi_6) = 0.47\; , \qquad
p(\pi_7) = 0.69 \; , \qquad p(\pi_8) = 0.31 \; .
\ee
This gives
\be
p(\pi_1) - p(\pi_5) = 0.07 \; , \qquad   
p(\pi_7) - p(\pi_3) = 0.06 \; .
\ee
Thus, while the planning paradox is clear, not all individuals follow it, 
justifying the probabilistic framework of QDT. Moreover, as is seen from 
the above equations, the difference between the compared prospect 
probabilities is rather small, lying on the boundary of statistical errors.

Concluding this section, the planning paradox has been explained away
by taking into account the impact of subjective beliefs and emotions 
in decision making via the attraction factor defined by
expression (\ref{eq20}). We stress also that the proposed framework
remains valid both for single-step as well as for multistep gambles.

\subsection{Discounting effects}

Generally, the term discounting addresses the problem of 
translating values from one time period to another. The larger the
discount rate, the more weight the decision maker places
on costs and benefits in the near term over costs and benefits
over the long term. Depending on the specification of the problem,
it is possible to distinguish several discounting effects, that 
we analyze in turn.

\subsubsection{Value discounting \label{jtbnwrtnv}}

According to classical utility theory, the costs and benefits
of an action can be evaluated by means of its 
utility, or its value to the decision maker. The benefits of an action
are, for instance, to receive an amount 
of money or any other useful object at a given time. When an 
action $x$ is made at time $t$, it has a utility $u(x,t)$. Assume, we start 
our analysis at time zero, $t=0$, when the action utility is $u(x,0)$. But 
the same action at a later time $t$ is $u(x,t)$, which may be different. The 
difference comes from the obvious understanding that what we get earlier we 
can start using earlier, hence, it is more useful than what we would get 
later, having less time for its use. A typical example is provided by the 
time value of money. An amount $x$ of money received at time $t = 0$ has a 
value $u(x,0)$. This money can bring a profit, increasing, after the period 
of time $t_n$ to the amount $x(1+r)^{t_n}$, where $r$ is an interest rate 
for a unit time interval. Therefore, the value of money $x$ today is larger 
than the value of the same amount of money after time $t_n$. Hence, 
it is natural to prefer $x$ now, instead of $x$ at a future time $t_n$.

In QDT, this preference for a receipt now rather than delayed can
be framed in the following decision making procedure. We consider
the intended actions of getting an amount of money now ($A_1$) 
or, the same amount, sometimes later ($A_2$). The different possible 
ways of using this money are described by a set $\{B_\mu\}$ of intended 
actions $B_\mu$. One makes a choice between the prospects  
\be
\label{eq51}
\pi_j = A_j B \; , \qquad 
B \equiv \bigcup_{\mu} B_\mu \qquad (j=1,2) \; .
\ee 
The prospect probabilities are 
\be
\label{eq52}
p(\pi_1) = \sum_\mu p(A_1 B_\mu) \; + \; q(\pi_1) \; , \qquad
p(\pi_2) = \sum_\mu p(A_2 B_\mu) \; + \; q(\pi_2) \; .
\ee
The fact that an amount of money now gives more possibilities than 
the same amount received later means that 
\be
\label{eq53}
\sum_\mu p(A_1 B_\mu) \; > \; \sum_\mu p(A_2 B_\mu) \; .
\ee
In addition, getting something later is more uncertain, hence, 
$q(\pi_1) > q(\pi_2)$. Then it is evident that $\pi_1 > \pi_2$. 

While the conclusion is the same as in classical utility theory,
what QDT brings additionally is the breakdown of the time value
into an objective component (the sums of probabilities in (\ref{eq53})
quantifying the investment and consumption opportunities) and
a subjective component $q(\pi)$ quantifying the emotional 
cost of various degrees of delaying.

\subsubsection{Event uncertainty \label{jtb2otq}}

Certain paradoxes arise because the problems are not
well-posed or are too ill-defined with some features remaining
unspecified or vague. Consider 
the typical example where one has to choose between 50 dollars 
now or a significantly larger amount, say 100 dollars, in a year. Proposing
a larger amount in the future is supposed to account for the 
discounting effect of the previous subsection. Indeed, given that
a given amount now is always preferred to the same amount 
in the future (assuming a normal growing economy), as 
explained in the previous subsection, one can expect to 
find some larger amount tomorrow that would be as attractive
as the proposed sum today. The ratio of the two sums defines
the discount factor of a given individual, which quantifies the value of
his/her time preference. The example comparing $\$ 50$ now to
$\$ 100$ in a year implicitly considers that the rational discount factor
cannot be less than $1/2$, or in other words, the interest rate
that would provide dividends to an investment of $\$ 50$ cannot be
larger than $100\%$, so that the sum of $\$ 100$ in a year should be more
attractive than the sum of $\$ 50$ received immediately. It turns out
that it is often observed that individuals prefer to get  $\$ 50$ now instead 
of $\$ 100$ in a year. This seems a priori quite puzzling.

In fact, there is no real mystery, even within classical
utility theory: because of the formulation of the 
problem, the related probabilities are not defined. And decision 
makers intuitively understand that the receipt of $\$ 50$ now is rather 
certain, while the sum of $\$ 100$ in a year is not certain at all. That 
is, one compares the lottery $L_1=\{ 0,0;\;\$ 50,1;\;\$ 100,0\}$ with  
the lottery $L_2=\{ 0,1-p;\;\$ 50,0;\;\$ 100,p\}$, where $p$ is not known.
It can be perceived to be small because of many reasons, e.g., lack of trust
in the commitment to deliver $\$ 100$ in a year due to uncertainties
associated with the possible death, bankruptcy or simply default
of the counter party, or uncertainty in the survival of the decision maker
who would not be in a position to enjoy the receipt of $\$ 100$ in a year.
Therefore, the expected utility of the first lottery is $U(L_1)=u(\$ 50)$,
while that of the second lottery is $U(L_2)=(1-p)u(0)+pu(\$ 100)$. For 
sufficiently small $p\ll 1$, it happens that $U(L_1)>U(L_2)$, justifying the 
preference of $L_1$ to $L_2$. The effect is referred to in the literature as 
``uncertainty aversion''.

In QDT, this effect is easily described in the same way as in subsection~\ref{jtbnwrtnv}. One compares the prospects of getting $\$ 50$ 
now ($\pi_1$) or $\$ 100$ in a year ($\pi_2$). The smaller probability of 
the second prospect implies inequality (\ref{eq53}). The process of 
waiting is related to anxiety \cite{37}, making the delayed event of 
getting money less attractive. And, by definition, the second prospect 
is less attractive since it is more uncertain. That is, $q(\pi_1)>q(\pi_2)$. 
The immediate result is that $\pi_1>\pi_2$. 

It is interesting to compare the two explanations. In classical utility 
theory, the preference for $\$ 50$ now instead of $\$ 100$ in a year
is accounted for by uncertainty aversion, translating into a small
subjective probability for the $\$ 100$ payoff to happen. In QDT, the
uncertainty aversion is embodied automatically into the attraction
factor $q(\pi)$, while the normal discounting effects associated
with different opportunities are included in the objective probabilities
$\sum_\mu p(A_j B_\mu)$.

\subsubsection{Preference reversal \label{gwrtbwl}}

A standard problem in classical
decision theory is revealed by a dynamic-inconsistency paradox
associated with the inversion of preferences, in which money
versus time preferences are inverted as the time horizon is changed.
To specify the problem, 
let us consider the following setup. There is a choice between $\$ 50$ 
now and $\$ 100$ in a year. As discussed in Sec.~\ref{jtb2otq} above, 
individuals almost always prefer $\$ 50$ now. But when there is a choice 
between $ \$50$ in ten years and $\$ 100$ in eleven years, human beings 
usually prefer $\$ 100$ in eleven years. This reversal of preference 
occurs notwithstanding the fact that the time difference between ten 
and eleven years is exactly the same as between zero and one, so that 
a pure rational discounting mechanism would predict the same consistent 
choice of the smaller amount at the earlier time. This reversal is 
usually associated with a trait characterizing human beings, called 
hyperbolic value discounting or generalized hyperbolic discounting \cite{38,39,40,41,42,42a}, such that the near events are characterized 
by larger discount rates than the events in a more distant future. 
The problem is that this explanation unavoidably leads to time inconsistency
since, when the decision maker reconsiders the same choice after ten 
years, he/she again would prefer $\$ 50$ today to $\$ 100$ in a year, thus 
again reversing the previous preference he/she expressed ten years
earlier.

The preference-reversal paradox finds a natural
explanation within QDT, since its formulation in terms of 
prospects implies that choices
considered at different times and planned for at
different future instants of time are actually different prospects, 
even though they are associated with equivalent actions. To be more precise, 
a correct definition of a prospect $\pi_j$ depends on the point in time 
$t_0$ when it is considered, as well as on the point in time $t$
for which it is planned to be realized. That is, strictly speaking, 
a prospect is a function $\pi_j(t,t_0)$. With this specification,
the above setup can be formalized as follows. 
Let the prospects of getting $\$ 50$ or $\$ 100$ 
correspond to the notations $\pi_1$ and $\pi_2$, respectively. At time 
$t_0=0$, there are four prospects. One is the prospect $\pi_1(0,0)$ of 
getting $\$ 50$ now. Another is the prospect $\pi_2(1,0)$ of getting 
$\$ 100$ in a year. The third prospect is $\pi_1(10,0)$ of getting 
$\$ 50$ in 10 years. And the fourth prospect $\pi_2(11,0)$ is 
getting $\$ 100$ in 11 years. As discussed above, the odds of 
getting $\$ 50$ now are more certain than those of getting $\$ 100$ 
in a year, hence 
\be
\pi_1(0,0) \; > \; \pi_2(1,0) \; . 
\ee
At the same time, both prospects of getting $\$ 50$ in ten years or 
$\$ 100$ in eleven years seem almost equally uncertain. However, the stake in 
the latter case is larger, which results in the preference 
\be
\pi_2(11,0) \; > \; \pi_1(10,0) \; .      
\ee

After time elapses to the decision at the point in time $t_0 = 10$, two 
new prospects become available. One is the prospect $\pi_1(10,10)$ of getting 
$\$ 50$ at this moment of time and another, $\pi_2(11,10)$ of getting 
$\$ 100$ one year later after $t_0=10$. Using the same arguments, 
one has 
\be
\pi_1(10,10) \; > \; \pi_2(11,10) \; .      
\ee
There is no contradiction between the above decisions, since 
different prospects are compared.

\section{Prospect dynamics}

\subsection{Definition of the discount factor}

The evolution of probabilities in classical decision theory are 
usually characterized by Markov equations \cite{36,43}. To determine
how the probability of a given prospect in QDT evolves as a function
of time, let us  consider a prospect $\pi_j(t,t_0)$ of deciding at time $t_0$ 
for the planned realization at a later time $t$. The corresponding prospect 
state is $|\pi_j(t,t_0)>$. Using the definitions of section \ref{jtwnaa},
the corresponding prospect operator is
\be
\label{eq54}
\hat P(\pi_j(t,t_0)) = | \pi_j(t,t_0) > < \pi_j(t,t_0) | \; .
\ee
The prospect probability is defined by the average (\ref{eq13}), which we denote
\be
\label{eq55}
p_j(t,t_0) \; \equiv \; < \hat P(\pi_j(t,t_0)) > \; .
\ee

We may assume that the mind strategy defined by Eq. (\ref{eq10}), which 
characterizes a given decision maker, does not change during the time during
which the decisions are made. In other words, the same decision maker is 
considered. Then, the prospect probability varies in time as 
\be
\label{eq56}
\frac{d}{dt} \; p_j(t,t_0) \; = \; < \frac{d}{dt} \;
\hat P(\pi_j(t,t_0)) > \; .
\ee

Let us define the decay rate $\al_j(t,t_0)$ of the prospect state 
$|\pi_j(t,t_0)>$ through the equation
\be
\label{eq57}
\frac{d}{dt} \; | \pi_j(t,t_0) > = -\; \al_j(t,t_0) \; | \pi_j(t,t_0) >.
\ee
The decay rate $\al_j(t,t_0)$ accounts for the possible disappearance
of opportunities as the future unfolds. The above equation is the definition 
of the decay rate. Since the latter depends on time, this definition does not 
necessarily imply that the time evolution of the prospect state is exponential.
And the following consideration will show that, really, there can occur 
different types of the time evolution.
 
Accomplishing the differentiation in the right-hand side of
Eq. (\ref{eq56}) yields
\be
\label{eq58}
\frac{d}{dt} \; p_j(t,t_0) \; = \; - \gm_j(t,t_0) p_j(t,t_0) \; ,
\ee 
where
\be
\label{eq59}
\gm_j(t,t_0) \equiv 2{\rm Re}[\al_j(t,t_0)] 
\ee
can be called the ``probability discount rate.''
Integrating equation (\ref{eq58}) gives the prospect probability
\be
\label{eq60}
p_j(t,t_0) = p_j(t_0,t_0) f_j(t,t_0) \; ,
\ee
with the discount factor
\be
\label{eq61}
f_j(t,t_0) \equiv \exp \left \{ 
- \int_{t_0}^t \gm_j(t',t_0)\; dt' \right \} \; ,
\ee
obeying the initial value condition $f_j(t_0,t_0)=1$. Equations (\ref{eq60}) 
and (\ref{eq61}) define the probability of a prospect, evaluated at an 
initial time $t_0$, which is to be realized at the instant of time 
$t$.

In the economic literature, the simplest and standard assumption is to 
assume a constant discount rate, corresponding to an exponential
discount factor. As reviewed by Cochrane \cite{Cochrane}, 
the exponential discount factor can be generalized into the concept of 
the stochastic discount factor which, by capturing the macro-economic 
risks underlying each security's value, provides a consistent pricing 
of all assets. Different models, such as the Capital Asset Pricing Model, 
multifactor models, term structure of bond yields, and option pricing 
can be derived as different specifications of the discount factor. 

\subsection{First-principle construction of discount rate}

Here in contrast, rather than deriving the form of the discount factor 
that corresponds to a specific economic model, we construct, by using 
general symmetry requirements, the possible generic functional dependencies
that the discount factor can take to describe the value of delayed 
payoffs. For this, we use the self-similar approximation theory 
\cite{44,45,46,47,48,49,50}. The idea is to start from an expansion of 
the discount rate valid for short time, that is believed to be generally 
valid. Then, particular conditions are implemented to construct
the functional forms that can be naturally associated with the initial 
expansion. The derivation of the corresponding discount factor proceeds 
through three successive steps. First, to improve the convergence property 
of a perturbative sequence, control functions, defined by an optimization 
procedure, are introduced. This idea forms the foundation of the optimized 
perturbation theory \cite{49,50}. The second pivotal idea is to consider 
the successive passage from one approximation to the next one as a dynamical
evolution on the manifold of approximants, which is formalized by the 
notion of group self-similarity. The third principal point is the 
introduction of control functions in the course of rearranging perturbative
asymptotic expansions by means of algebraic transforms. We use the variant 
of the self-similar approximation theory \cite{44,45,46,47,48,49,50} 
employing the self-similar factor approximants \cite{51,52,53,54,55},
based on the property that the control parameters entering
the self-similar factors can be completely defined from a given 
asymptotic expansion by the so-called accuracy-through-order matching 
method. This approach was shown to be essentially more accurate than 
the method of Pad\'e approximants \cite{56}. Moreover, the latter method,
as is well known, does not allow a unique reconstruction of the sought function, 
but results in a whole table of approximants for each given approximation order. 
Contrary to this, the factor approximants are uniquely defined. In addition 
to providing reconstruction with a very good accuracy of rational functions, as 
the Pad\'e method does, the method of factor approximants determines 
irrational and transcendental functions with excellent precision \cite{51,52,53,54,55}. 
These approximants also allow one to reconstruct a wide class of functions exactly.

In its applications to the construction of the functional dependence
of the discount factor, we proceed as follows.
First, we note that, in full generality, the probability discount 
rate $\gm_j(t,t_0)$ can be positive as well as 
negative. This is because the prospect probabilities are normalized 
according to condition (\ref{eq15}). Consequently, if there are 
diminishing probabilities, then there should exist increasing 
probabilities in order that normalization (\ref{eq15}) be always 
valid. For instance, if the probability of getting something attractive, 
like money, diminishes with time, then the probability of getting 
nothing, respectively, increases. Therefore, in what follows, it 
is sufficient to consider only decreasing probabilities, related to 
getting something appealing,  keeping in mind that there exist as well 
their increasing counterparts defined through the normalization (\ref{eq15}).     
The condition, that the probability discount rate $\gm_j(t,t_0)$ is a nonincreasing 
function of time, reads
\be
\label{eq62}       
\frac{d}{dt} \; \gm_j(t,t_0) \; \leq 0 \; .
\ee

To go further, we assume that the rate $\gm_j(t,t_0)$ is an analytic function 
of $t$ in the vicinity of the initial time $t = t_0$. This means that 
the expansion
\be
\label{eq63}
\gm_j(t,t_0) \simeq \gm_j \; \sum_{n=0}^k a_n (t -t_0)^n \; ,
\ee
where $\gm_j\equiv\gm_j(t_0,t_0)$ is the spot rate and $a_0=1$, is valid for 
asymptotically small $t-t_0\ra 0$. The upper limit $k$ of the summation 
can be taken to infinity.

Then, the method of self-similar factor 
approximants \cite{51,52,53,54,55} mentioned above is used
to construct the general class of functions corresponding
to the expansion (\ref{eq63}). This amounts to
extrapolate the asymptotic series (\ref{eq63}), valid for 
small $t-t_0$, to the region of all $t > t_0$. 
Extrapolating, by means of the 
self-similar factor approximants \cite{51,52,53,54,55}, the asymptotic series 
(\ref{eq63}) under condition (\ref{eq62}) gives 
\be
\label{eq64}
\gm_j(t,t_0) = \gm_j \left ( 1 + \frac{t-t_0}{t_j}
\right )^{-n_j} \; ,
\ee
where $t_j$ is a time scale and $n_j\geq 0$. We stress the non-trivial
nature of the construction of the function (\ref{eq64}) by
the self-similar factor approximants, which makes appear
the exponent $n_j$. This exponent plays a key role in 
structuring the form of the discount factor.

\subsection{Four classes of discount factors}

Four types of discount factors are predicted, corresponding
to the four different sets: (i) $n_j=0$; (ii)  $0< n_j<1$; (iii) $n_j=1$; 
and (iv) $1<n_j$.
\vskip 2mm

(i) $n_j = 0$. The discounting function (\ref{eq61}) is the 
simple exponential
\be
\label{eq65}
f_j(t,t_0) =\exp \{ - \gm_j (t-t_0) \} \; .
\ee
This type of discount factor is standard in the value-discounting problems.
We may notice that reparametrizing Eq. (\ref{eq65}) with the 
substitution $\dlt\equiv\exp(-\gm_j)$ yields an equivalent expression
$f_j(t,t_0) = \dlt_j^{t-t_0}$.

\vskip 2mm

(ii) $0< n_j< 1$. The discounting function (\ref{eq61}) takes the form
\be
\label{eq67}
f_j(t,t_0) = \exp\left \{ - \; 
\frac{\gm_j t_j}{1-n_j} \; \left [ \left ( 1 +
\frac{t-t_0}{t_j} \right )^{1-n_j} \; - \; 1 
\right ] \right \} \; .
\ee
At short times $t-t_0 < t_j$, the expression $f_j(t,t_0)$ reduces 
approximately to the pure exponential form. However, for large times, such 
that $t\gg t_0, t_j$, this $f_j(t,t_0)$ is approximated by the function
\be
\label{e168}
f_j(t,t_0) \simeq \exp\left \{ -\; \frac{\gm_j t_j}{1-n_j} 
\left ( \frac{t}{t_j} \right )^{1-n_j} \right \} \; ,
\ee
called the stretched exponential (see, e.g., Chapter 6 of 
Ref.~\cite{SornetteCritical}). Stretched exponential relaxation of a 
macroscopic variable to an equilibrium is well-known in physics, such 
as in ``complex'' fluids \cite{Klinger}, glasses
\cite{glass,glass2,glass3,Phillips,Palmer}, porous media, semiconductors, 
etc., a law known under the name Kohlrausch--Williams--Watts law
\cite{glass,Phillips}. The stretched-exponential decay of the discount 
factor as a function of time reflects a decay slower than exponential 
of the time value of future payoffs. An even slower decay is found
for the next case.

\vskip 2mm

(iii) $n_j=1$. The discounting function (\ref{eq61}) reads
\be
\label{eq69}
f_j(t,t_0) = \frac{1}{[1+(t-t_0)/t_j]^{\gm_j t_j} } \; .
\ee
This recovers the postulated form associated with so-called
generalized hyperbolic discounting or, simply, 
hyperbolic discounting function \cite{38,39,40,41,42}, which 
seems to account better for the observed time-preference of human beings
than the standard exponential form (\ref{eq65}).

\vskip 2mm

(iv) $n_j>1$. Eq. (\ref{eq61}) leads to
\be
\label{eq70}
f_j(t,t_0) = \exp \left \{ - \; \frac{\gm_j t_j}{n_j-1} \;
\left [ 1 \; - \; \frac{1}{(1+(t-t_0)/t_j)^{n_j-1} } \right ]
\right \} \; .
\ee
At short times, $f_j(t,t_0)$ is again well-approximated
by an exponential form. However, at large times, when $t-t_0 \gg t_j$, 
the factor $f_j(t,t_0)$ tends to a non-zero limit
\be
\label{hyneth}
\lim_{t\ra\infty} f_j(t,t_0) = 
\exp\left ( - \; \frac{\gm_j t_j}{n_j-1} \right ) \; .
\ee
This is in contrast with the previous cases (\ref{eq65}) to (\ref{eq69}) 
and with the standard assumption that
$f_j(t,t_0)$ tends to zero at large times because individuals
do not care for events that are very-very far in the future.
This new regime is a priori unexpected and surprising, because it implies
that payoffs or costs that are very far in the future still contribute
a finite amount to the likelihood of a given prospect. In common
terms, according to (\ref{eq70}) leading to (\ref{hyneth}),
extremely far ahead outcomes are not discounted to zero, but
provide a finite input to the effective utility of the decision maker.
While providing perhaps the most dramatic rupture with standard
discounting and decision making theory, we believe that the form 
(\ref{eq70}) leading to the bizarre result (\ref{hyneth}) is actually
formalizing an important element of decision making. Specifically,
very low to zero discount rates are presently being discussed for 
analyzing intergenerational public policy choices \cite{Weitzmann,
Bazelon,Rambaud}. These policies encompass issues such as global 
warming and nuclear waste disposal. Nuclear waste disposal, in particular, 
involves time scales up to millions of years over which mankind will have
to continue to monitor and watch the long-lived radionuclides resulting
from the burning of nuclear fuel in nuclear plants. The ongoing challenge 
is to characterize distant future costs or benefits in a way that is 
relevant for policy makers, who must evaluate trade-offs today.

\subsection{Prospect-dependent discount rates}

In full generality, different intended actions can be characterized 
by different discount functions. Even if, for simplicity, the same 
discount function is employed, then different actions can have 
different decay rates $\gm_j$ or different time scales $t_j$. This 
can lead to a reversal of natural preferences. 

For example, let a prospect $\pi_1$ be preferred to $\pi_2$, if they 
are realized at the initial time $t_0$, so that for their probabilities 
the following inequality 
holds:
\be
\label{eq71}
\frac{p_1(t_0,t_0)}{p_2(t_0,t_0)} \; > \; 1 \; .
\ee
But, if these prospects are planned to be realized at a later time 
$t$, then their probabilities form the ratio
\be
\label{eq72}
\frac{p_1(t,t_0)}{p_2(t,t_0)} = 
\frac{p_1(t_0,t_0) f_1(t,t_0)}{p_2(t_0,t_0) f_2(t,t_0)} \; .
\ee
It may happen that at some moment of time $t_{rev}$, their 
probabilities reverse, so that for $t>t_{rev}$,
\be
\label{eq73}
\frac{p_1(t,t_0)}{p_2(t,t_0)} \; < \; 1 \qquad 
( t > t_{rev} ) \; ,
\ee
which implies preference reversal. This phenomenon, known as ``time 
inconsistency'' in the literature, is usually associated with 
non-exponential discount factors. It may also occur with exponential 
discount factors, when the discount rate is different from the 
risk-adjusted return on saving (see, e.g., Chapter 15 in Ref.~\cite{6}). 
Within QDT, time reversal can also occur for the exponential discount 
factor when the discount rates of two prospects are different. The 
reversal time in the case of the exponential discounting (\ref{eq65}) 
is
\be
\label{eq74}  
t_{rev} =  t_0 + \frac{1}{\gm_1 -\gm_2} \;
\ln \; \frac{p_1(t_0,t_0)}{p_2(t_0,t_0)}  \; ,
\ee
which exists for $\gm_1>\gm_2$. In the case of the hyperbolic 
discounting (\ref{eq69}), with $\gm_j t_j=1$, the reversal time reads 
as
\be
\label{eq75}
t_{rev} = t_0 + 
\frac{p_1(t_0,t_0)-p_2(t_0,t_0)}
{\gm_1 p_2(t_0,t_0)-\gm_2 p_1(t_0,t_0)} \; ,
\ee
which exists under the condition
\be
\frac{\gm_1}{\gm_2} \; > \; 
\frac{p_1(t_0,t_0)}{p_2(t_0,t_0)} \; .
\ee
Recall that the initial time $t_0$ corresponds to the 
planning time when the decision maker evaluates a prospect 
that is assumed to be realized at the point in time $t\geq t_0$. 
Thus, the planning time $t_0$ is also a variable, which shifts 
when the decision-maker re-evaluates his/her plans. As a consequence,
there is no preference-reversal paradox within QDT, as explained in 
Sec.~\ref{gwrtbwl}.

\section{Conclusion}

We have presented a novel approach to decision making, based 
on the mathematical techniques of complex Hilbert spaces over a 
lattice of composite prospects. Such techniques are typical for the 
theory of quantum measurements, which explains the name 
``Quantum Decision Theory'' (QDT). We stress that this does not presuppose 
that decision makers are assumed to be quantum objects. The employed 
mathematical methods are just the most convenient tool for taking into 
account such notions as risk and uncertainty which have strong
emotional effects in decision making. QDT makes it possible to explain the 
paradoxes appearing in the application of classical utility theory to 
decision making. In the present paper, we have analyzed the stylized effects 
and paradoxes, associated with dynamic aspects of decision theory, such as time 
inconsistency, planning paradox, value discounting, event uncertainty, and 
preference reversal. These temporal effects have not been considered in our 
previous articles on QDT \cite{15,16,22} and the treatment offered here is 
original. We have also suggested a constructive approach for deriving 
the evolution equations for the prospect probabilities. The derived 
discount functions provide a novel classification of possible discount factors,
which include the previously known cases (exponential or hyperbolic discounting),
but also predicts a novel class of discount factors that can be applied
for very long-term discounting situations.

One of the basic conclusions of QDT is the necessity of taking into account 
not merely the utility of the considered prospects, as in classical utility 
theory, but also the attractiveness of the related alternatives. This is accounted 
for by the attraction factor, whose appearance is due to the use of the quantum 
techniques. The attraction factor characterizes the level of attractiveness of each 
prospect with regard to the risk and uncertainty associated with the choice among 
the related alternatives. In that way, the attraction factor is a new measure of risk in 
decision making. Mathematically, its appearance is caused by the use of quantum
rules in defining the prospect probabilities. And its meaning is the 
characterization of the perceived level of risk associated with emotions and subconscious 
processes that influence decision making. In brief, we can say that the physics of risk in decision making, 
described by the attraction factor, embodies the existence of subconscious feelings, 
emotions, and biases.

The notion of risk is met in many applications,
such as economics, finance, psychology, and so on. In all these applications,
it is always connected with the process of taking decisions. Therefore, to elucidate 
the physics of risk, one needs, first of all, to understand its meaning in decision 
making. Without such an understanding, it is impossible to properly employ this 
notion in applications to other fields. As we have shown, the evaluation of risk 
presents two sides. In addition to a first contribution, measured, e.g.,
through the risk-aversion coefficients \cite{4,5} and the lottery dispersion, it 
is necessary to take into account its subjective part caused by emotions. A
principal result of our theory is that, despite the subjectivity of 
the emotional side of risk, it is possible to naturally take it into account  
in a logical and mathematically self-consistent way. Our QDT is the first 
mathematically rigorous realization of the old Bohr idea \cite{Bohr}  that mental human 
processes can be described by techniques of quantum theory.

Obviously, taking correct decisions is of paramount 
importance. This is why the developed theory can find numerous applications. 
Several illustrations have been analyzed in the present paper. We have concentrated 
our attention here on temporal effects, related to time inconsistency, which have not been 
considered in our previous articles.

\vskip 5mm

{\bf Acknowledgements}

\vskip 2mm

We are grateful to F. Schweitzer for the possibility of presenting our 
results in the International Workshop on the ``Physics Approach to Risk'', 
Z\"urich, Switzerland, October, 27-29 2008. We also thank the participants 
of the seminar on ``Modeling Socio-Economic Systems and Crises'', 
organized by the ETH Competence Center for Coping with Crises 
in Socio-Economic Systems, for their constructive remarks. We acknowledge financial support 
from the ETH Competence Center ``Coping with Crises in Complex 
Socio-Economic Systems" (CCSS) through ETH Research 
Grant CH1-01-08-2. We appreciate
discussions with E.P. Yukalova.

\end{document}